\documentclass[epj,twocolumn]{webofc}
\usepackage[varg]{txfonts}   % Web of Conferences font
\woctitle{Hadron Collider Physics symposium 2012}

\newcommand{\GeV}{\ensuremath{\rm\ GeV}} 
\newcommand{\TeV}{\ensuremath{\rm\ TeV}} 

\newcommand{\fbinv}{\ensuremath{\rm\ fb^{-1}}}
\newcommand{\Lint}{\ensuremath{\rm\ L}}
\newcommand{\pb}{\ensuremath{\rm\ pb}}
\newcommand{\stat}{\ensuremath{\rm\ (stat.) \, }}
\newcommand{\syst}{\ensuremath{\rm\ (syst.) \,  }}
\newcommand{\statsyst}{\ensuremath{\rm\ (stat.+syst.) \, }}
\newcommand{\lumi}{\ensuremath{\rm\ (lum.) \,  }}

\newcommand{\scale}{\ensuremath{\rm\ (scale)  \, }}
\newcommand{\PDF}{\ensuremath{\rm\ (PDF)  \, }}
\newcommand{\ttbar}{\ensuremath{{t\bar{t}}}}

\newcommand{\BR}{\ensuremath{ {\rm BR}  }}

\begin{document}

\title{Measurements of the Top Quark Pair-Production Cross Section}

\author{
Frank-Peter Schilling\inst{1}\fnsep\thanks{\email{fpschill@cern.ch}}
(on behalf of the ATLAS, CDF, CMS, D0 collaborations) 
}

\institute{
Karlsruhe Institute of Technology (KIT), Karlsruhe, Germany 
}

\abstract{
Measurements of the inclusive and differential cross section for the production of top quark pairs in proton-(anti)proton collision at center-of-mass energies of 1.96, 7.0 and 8.0 TeV are presented and compared with the latest theory predictions and Monte-Carlo models. In addition, first measurements of the production of top quark pairs in association with additional jets or with a boson are highlighted. All measurements are in good agreement with the Standard Model.
}

\maketitle

%%%%%%%%%%%%%%%%%%%%%%%%%%%%%%%%%%%%%%%%%%%%%%%%%%%%%%%%%%%%%%%%%%%%%%%%%%%%%%%

\section{Introduction}
\label{intro}

The top quark is the heaviest known fundamental particle. Due to its Yukawa coupling being close to unity, it is often believed to play a special role in the electroweak symmetry breaking. Its presence in virtual loops relates the top quark mass $m_t$ with the one of the Higgs boson. Many models for new physics beyond the Standard Model (SM) predict new particles which decay predominantly into top quarks (e.g. Super-Symmetry, SUSY), and new particles may also be produced in top decays. For these reasons, it is important to study the top quark in great detail, in order to check the consistency of the SM, and to look for contributions from new physics.

The production of top quarks in hadron collisions has been studied extensively at the $p\bar{p}$ collider Tevatron (CDF and D0 experiments) at a center-of-mass energy of $\sqrt{s} \leq 1.96 \rm\ TeV$ (see e.g.~\cite{Deliot:2010ey}), and more recently at the LHC with $\sqrt{s} = 7$ and $8 \rm\ TeV$ (see e.g.~\cite{topreview}), using the ATLAS and CMS detectors. The focus of this article lies with the latest experimental measurements of the inclusive and differential cross section for the production of $t\bar{t}$ pairs. Other properties of the top quark and its interactions are discussed in Refs.~\cite{talk-palencia,chiarelli-talk,shabalina-talk,stelzer-talk,demina-talk}, while the theory status is presented in Refs.~\cite{talk-mitov,kamenik-talk}.

\section{Total Cross Section}

The total cross section for the production of top quark pairs, $\sigma_{t\bar{t}}$, has been calculated to approximate next-to-next-to-leading (NNLO) order by various groups. Recently, most of the contributions to the exact NNLO result have been calculated, notably the $q\bar{q}$ initial state~\cite{Baernreuther:2012ws}, which is dominating at the Tevatron (see Ref.~\cite{talk-mitov} for details). The $gg$ initial state, which is dominant at LHC, remains the last missing ingredient.

Measurements of $\sigma_{t\bar{t}}$ have been performed in essentially all decay modes, which are classified according to the decay of the W-bosons: lepton+jets ($\ttbar \rightarrow W^+b W^-\bar{b} \rightarrow q \bar{q}' b l \bar{\nu}_{l} \bar{b} + \bar{l} \nu_l b q \bar{q}' \bar{b}$), 
di-lepton ($\ttbar \rightarrow W^+b W^-\bar{b} \rightarrow \bar{l} \nu_l b l' \bar{\nu}_{l'} \bar{b}$)
and hadronic ($\ttbar \rightarrow W^+b W^-\bar{b} \rightarrow q \bar{q}' b q'' \bar{q}''' \bar{b}$).
The lepton+jets mode has a large branching fraction and moderate background (mostly W+jets), the di-lepton mode is very clean but has a small branching fraction, while the hadronic mode suffers from huge QCD multi-jet background.

The cross section measurements, which traditionally were performed by means of a simple counting experiment, today frequently employ sophisticated likelihood fits to one or more discriminating distributions, often incorporating the effects of systematic uncertainties via nuisance parameters to achieve ultimate precision. The dominating systematic uncertainties are typically due to the imperfect knowledge of the jet energy scale and b-tagging efficiency, as well as due to the modeling of signal and backgrounds in the Monte-Carlo (MC) simulation.

\subsection{Tevatron}

The most precise individual measurements of $\sigma_{t\bar{t}}$ at the Tevatron have been performed in the lepton+jets channel. D0 has measured
$\sigma_{t\bar{t}} = 7.78 ^{+0.77}_{-0.64} \pb$ (precision $9.1\%$)~\cite{Abazov:2011mi}, using a dataset corresponding to an integrated luminosity of $\Lint =5.3 \fbinv$ and employing a profile likelihood fit to the two-dimensional distribution of jet and b-tag multiplicity. The best single channel result from CDF, $\sigma_{t\bar{t}} = 7.70 \pm 0.52 \pb$ (precision $6.8\%$)~\cite{Aaltonen:2010ic}, was obtained using $\Lint =4.6\fbinv$ of data and based on a Neural Network discriminant as well as combining measurements without and with the use of b-tagging.

Recent measurements have been extended to include the full Tevatron dataset. In particular, CDF has measured $\sigma_{t\bar{t}} = 7.47 \pm 0.50 \stat \pm 0.53 \syst \pm 0.46 \lumi \pb$ in the di-lepton channel ($\Lint = 8.8 \fbinv$, precision 11.5\%)~\cite{cdf10878}. Another new measurement by CDF in the low statistics $\tau$ plus $e/\mu$ plus jets channel reports $\sigma_{t\bar{t}} = 8.2 \pm 2.3 \stat ^{+1.2}_{-1.1} \syst \pm 0.5 \lumi \pb$, based on 36 selected events in $\Lint = 9.0 \fbinv$~\cite{cdf10562}.

\begin{figure}
%\sidecaption
\centering
\includegraphics[width=0.99\linewidth]{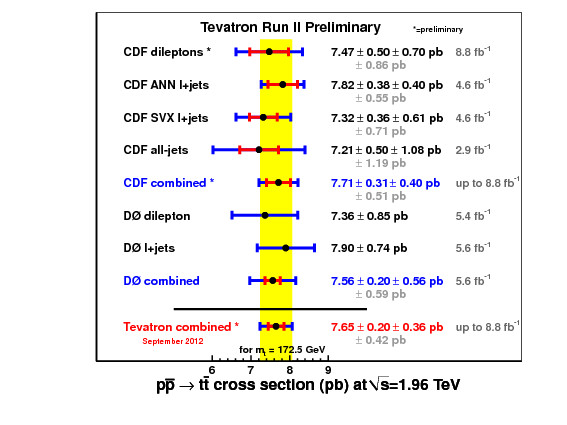}
\caption{Individual measurements of $\sigma_{t\bar{t}}$ at the Tevatron, as well as their combination~\cite{tevxscomb}.}
\label{fig:tevsummary}
\end{figure}

CDF and D0 recently combined for the first time their $\sigma_{t\bar{t}}$ measurements, based on a careful treatment of correlated systematic uncertainties (see Fig.~\ref{fig:tevsummary}). The result is $\sigma_{t\bar{t}} = 7.65 \pm 0.20 \stat \pm 0.36 \syst \pb$ (precision 5.5\%)~\cite{tevxscomb}, in good agreement with the theory prediction of $7.24 ^{+0.15}_{-0.24} \scale  ^{+0.18}_{-0.12} \PDF $ (for $m_t=172.5 \GeV$) which includes the exact NNLO $q\bar{q} \rightarrow \ttbar$ contribution ~\cite{talk-mitov}.

\subsection{LHC at 7 TeV}

Both ATLAS and CMS have performed a suite of measurements of $\sigma_{t\bar{t}}$ in many different channels at $\sqrt{s}= 7 \TeV$, using the 2010-2011 dataset. In the lepton+jets channel, ATLAS reported $\sigma_{t\bar{t}}= 179.0 \pm 9.8 \statsyst \pm 6.6 \lumi \pb$ (precision 7\%)~\cite{ATLAS-CONF-2011-121} using $\Lint = 0.7 \fbinv$ of data and performing  a likelihood fit based on various kinematic variables. Similarly, CMS obtained $\sigma_{t\bar{t}}= 158.1 \pm 2.1 \stat \pm 10.2 \syst \pm 3.5 \lumi \pb$ (precision 7\%, $\Lint = 2.3 \fbinv$)~\cite{:2012cj} from a fit to the secondary vertex mass distribution in bins of jet and b-tag multiplicity. In both cases, effects from systematic uncertainties were incorporated into the likelihood.

Recently, ATLAS has performed a measurement in the lepton+jets channel which is based on soft muon tagging to identify b-jets, which leads to corresponding systematic uncertainties which are largely orthogonal to the ones when using impact parameter based methods. The result is $\sigma_{t\bar{t}}= 165 \pm 2 \stat \pm 17 \syst \pm 3 \lumi \pb$ (precision 11\%, $\Lint=4.7 \fbinv$)~\cite{ATLAS-CONF-2012-131}.

CMS has presented a new measurement in the di-lepton channel based on $\Lint = 2.3 \fbinv$ of data and using a profile likelihood fit which incorporates systematic uncertainties. The result, $\sigma_{t\bar{t}}= 161.9 \pm 2.5 \stat ^{+5.1}_{-5.0} \syst \pm 3.6 \lumi \pb$ (precision 5\%)~\cite{:2012bta} is the most precise measurement of this quantity to date.

Other, although less precise, measurements were performed by both collaborations in the tau+jets~\cite{cmspas-top-11-004,Aad:2012vna} and hadronic~\cite{ATLAS-CONF-2012-031,CMS-PAS-TOP-11-007} channels.

\begin{figure}
\centering
\includegraphics[width=0.99\linewidth]{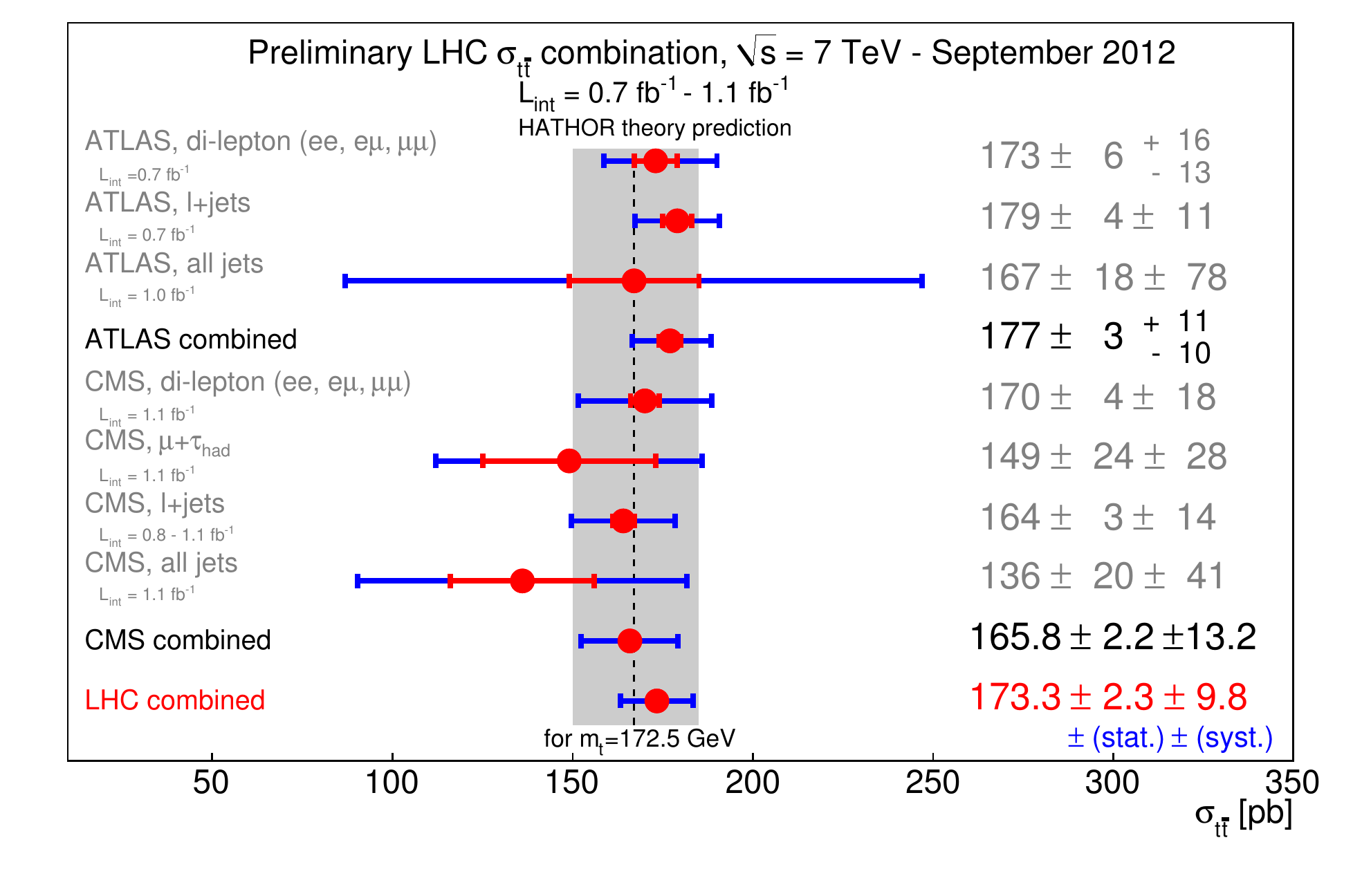}
\caption{Measurements of $\sigma_{t\bar{t}}$ at the $\sqrt{s}= 7 \TeV$ LHC, including their combination~\cite{lhc7combi}.}
\label{fig:lhc7summary}
\end{figure}

Recently, ATLAS and CMS presented a first combined measurement of $\sigma_{t\bar{t}}$, using existing per-experiment combinations as inputs. The result, shown in Fig.~\ref{fig:lhc7summary}, is $\sigma_{t\bar{t}}= 173.3 \pm 2.3 \stat \pm 9.8 \syst \pb$~\cite{lhc7combi}. Given that the combination does not yet include the very latest measurements, e.g. the precise CMS di-lepton result~\cite{:2012bta}, the precision of the combined result is 5.8\%. Again, the measurements are in very good agreement with most precise theory predictions.

\begin{figure}
\centering
\includegraphics[width=0.99\linewidth]{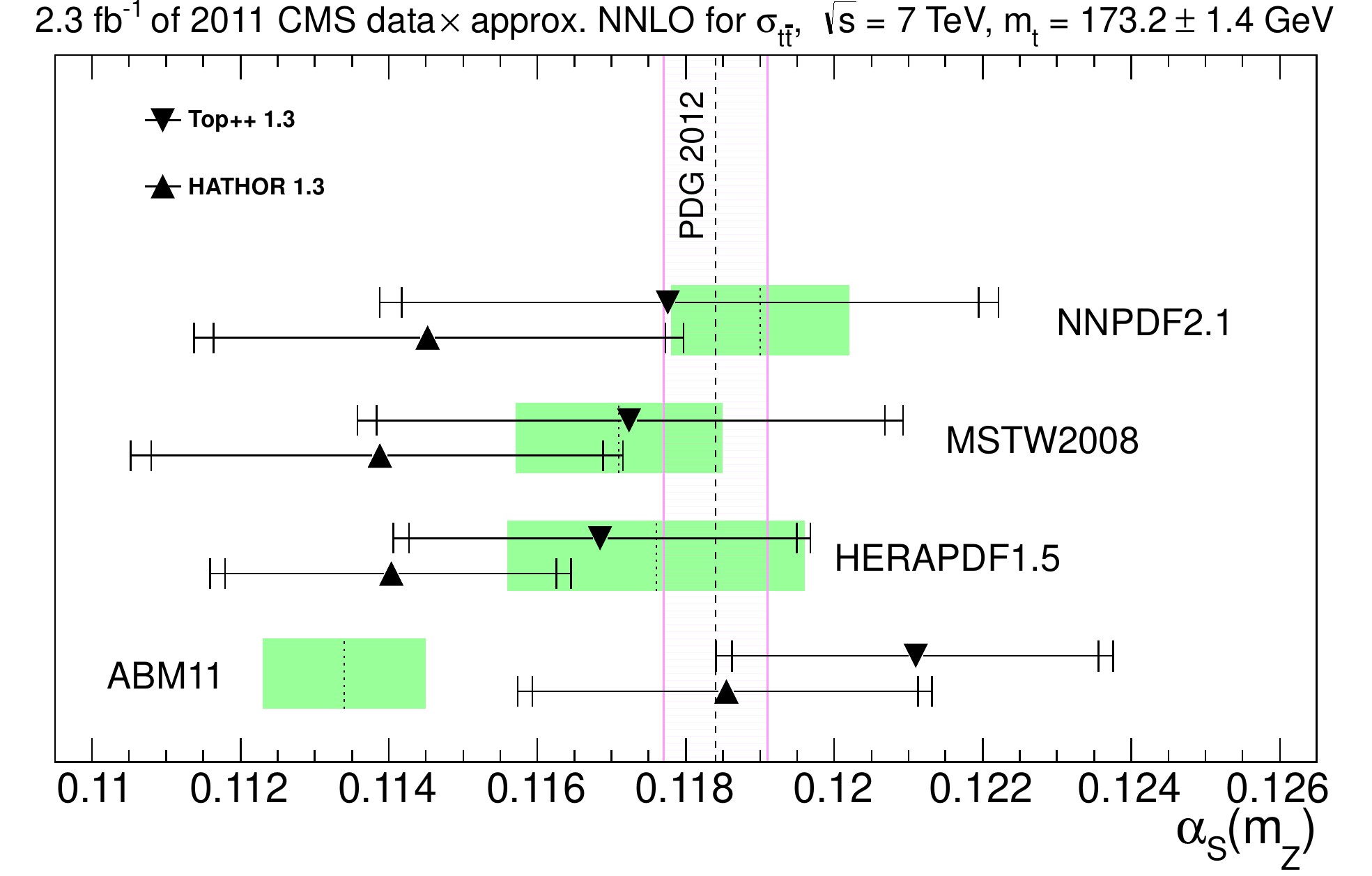}
\caption{Determination of $\alpha_s$ from the $t\bar{t}$ cross section by CMS, for two theory calculations and four different PDF sets~\cite{CMS-PAS-TOP-12-022}.}
\label{fig:alphas}
\end{figure}

For a given set of parton density functions (PDF), the theory value of $\sigma_{t\bar{t}}$ depends on the strong coupling $\alpha_s$ and $m_t$. This can be used to extract the value of $\alpha_s$, given the experimental cross section and mass measurements. CMS has performed such a determination of $\alpha_s$ using their di-lepton measurement~\cite{:2012bta} and the world average $m_t$ value, as well as for two different approximate NNLO theory calculations and four different sets of PDF. The result~\cite{CMS-PAS-TOP-12-022}, shown in Fig.~\ref{fig:alphas}, corresponds to the first $\alpha_s$ determination from top quark events. Its precision is already comparable to the one obtained using jet rates in hadron collisions. The lower $\alpha_s$ values obtained with the theory calculation obtained with HATHOR~\cite{Aliev:2010zk,Moch:2012mk} compared with the one using TOP++~\cite{Czakon:2011xx,Baernreuther:2012ws,Czakon:2012zr} are due to the larger value of the predicted $\sigma_\ttbar$.

\subsection{LHC at 8 TeV}

\begin{figure}
\centering
\includegraphics[width=0.99\linewidth]{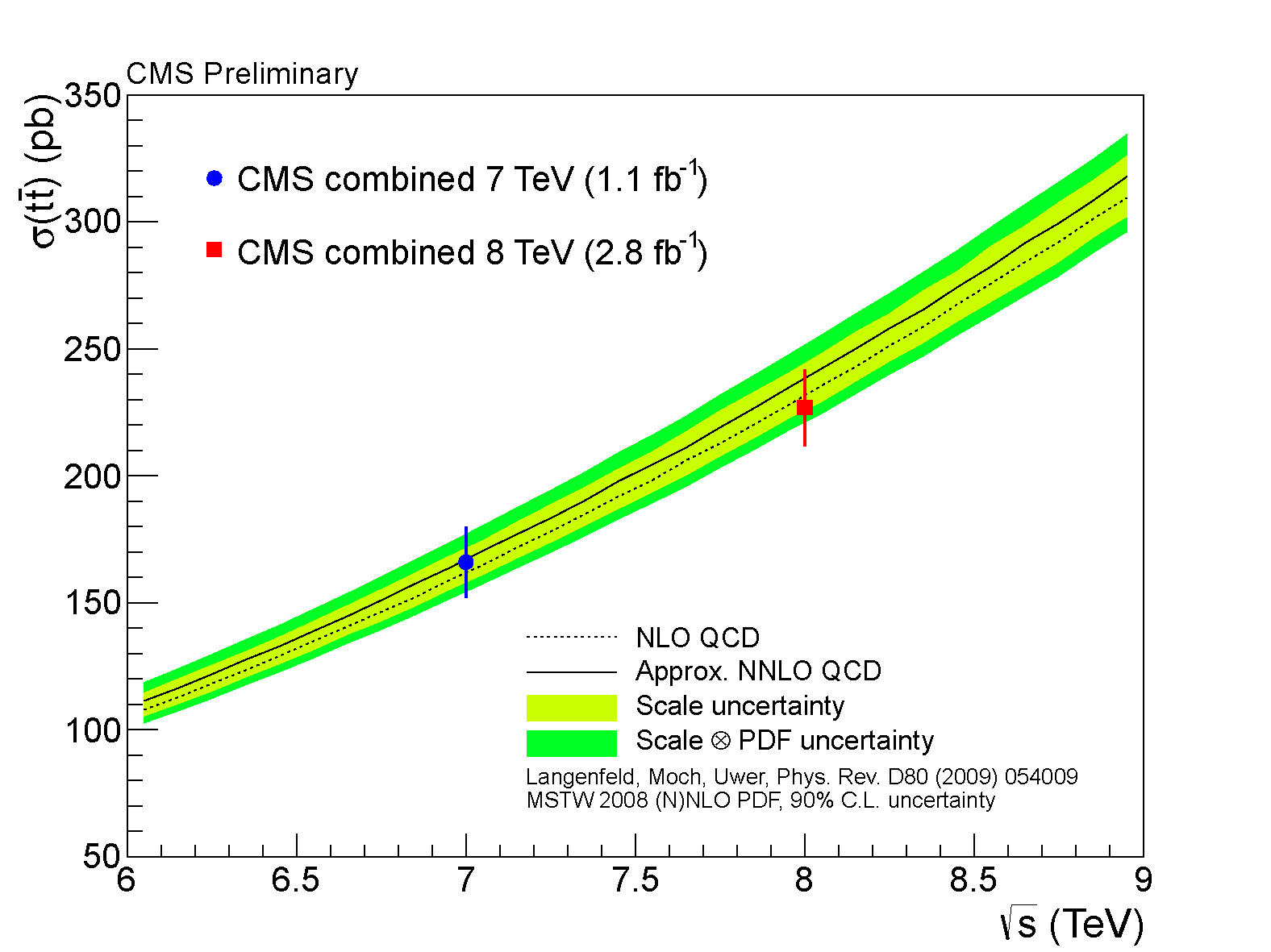}
\caption{Comparison of CMS $\sigma_{t\bar{t}}$ measurements at $\sqrt{s}=7$ and $8 \TeV$ with approximate NNLO theory~\cite{CMS-PAS-TOP-12-006,CMS-PAS-TOP-12-007}.}
\label{fig:lhcvscms}
\end{figure}

The energy dependence of $\sigma_{t\bar{t}}$ has been verified by measurements at the increased LHC center-of-mass energy of $\sqrt{s}= 8 \TeV$ during the 2012 run. CMS performed first measurements in both the lepton+jets and di-lepton channels based on $\Lint = 2.8 \fbinv$ of data, which yield a combined result of  $\sigma_{t\bar{t}}= 227 \pm 3 \stat \pm 11 \syst \pm 10 \lumi \pb$~\cite{CMS-PAS-TOP-12-006,CMS-PAS-TOP-12-007}. Fig.~\ref{fig:lhcvscms} shows the CMS measurements at 7 and 8 TeV, compared with theory.

More recently, ATLAS presented a measurement in the lepton+jets channel, based on a data sample corresponding to $\Lint = 5.8 \fbinv$. The result is $\sigma_{t\bar{t}}= 241 \pm 2 \stat \pm 31 \syst \pm 9 \lumi \pb$~\cite{ATLAS-CONF-2012-149}. Both results are in good agreement with the  theory value of $220 ^{+13}_{-11} \scale ^{+5}_{-6} \PDF \pb$~\cite{Czakon:2012pz}.

\section{Differential Cross Sections}

Besides measurements of the inclusive $\ttbar$ cross section, the abundance of $\ttbar$ pairs produced at LHC also allows for detailed studies of differential distributions. Besides comparisons with theory calculations and MC models (such as MC@NLO, POWHEG, MADGRAPH, ALPGEN or SHERPA), which allow to reduce current systematic uncertainties associated with the modeling of $\ttbar$ production, differential measurements are also vital for new physics searches where SM top quark production is a large background, in particular Higgs boson production and SUSY searches.

Experimentally, the measured differential distributions are unfolded to the level of stable hadrons or to the parton level, in order to facilitate comparisons with theory and across experiments. The unfolded cross sections are quoted either within the visible phase space corresponding to the kinematic selection applied for the measurement, or extrapolated to the full phase space.

\begin{figure}
\centering
\includegraphics[width=0.9\linewidth]{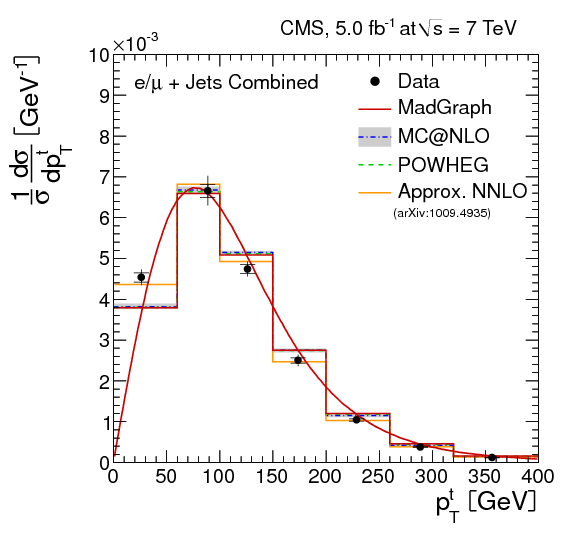}
\caption{Normalized top quark $p_T$ distribution as measured by CMS~\cite{:2012qka}, compared with various MC models as well as with an approximate NNLO theory calculation~\protect\cite{Kidonakis:2010dk}.}
\label{fig:pttop}
\end{figure}

CMS has measured the normalized transverse momentum ($p_T$) distribution of the top quark in the lepton+jets channel based on $5.0\fbinv$ of data~\cite{:2012qka}, see Fig.~\ref{fig:pttop}. While the data exhibit a  somewhat softer $p_T$-dependence compared with the MADGRAPH, POWHEG and MC@NLO models, the data are better reproduced by an approximate NNLO QCD calculation~\cite{Kidonakis:2010dk}. A similar trend was observed by an earlier D0 measurement~\cite{Abazov:2010js}.

Measurements of the normalized top quark rapidity, $y_t$, the $p_T$ and the rapidity of the $\ttbar$ system, as well as of the missing transverse energy distribution in the lepton+jets channel were all found to be in good agreement with current Monte-Carlo models~\cite{:2012qka,CMS-PAS-TOP-12-019,Aad:2012hg}.

\begin{figure}
%\sidecaption
\centering
\includegraphics[width=0.9\linewidth]{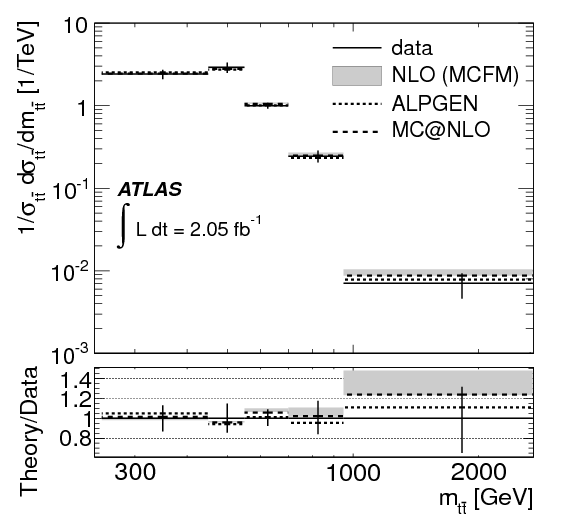}
\caption{Normalized cross section as a function of  $M_\ttbar$ as measured by ATLAS, compared with theory calculations and MC models~\cite{Aad:2012hg}.}
\label{fig:mttbar}
\end{figure}

The distribution of the top quark pair invariant mass, $M_\ttbar$, is sensitive to contributions due to the production of new, heavy particles which decay into $\ttbar$ pairs. While dedicated "bump-hunt" searches for resonances which decay into $\ttbar$ are not discussed here (see Ref.~\cite{stelzer-talk} instead), unfolded measurements of the normalized $M_\ttbar$ distribution have been performed by both ATLAS~\cite{Aad:2012hg} (see Fig.~\ref{fig:mttbar}) and CMS~\cite{:2012qka} covering the range up to $M_\ttbar \sim 2 \TeV$. Again, good agreement is observed with current MC models as well as with theory calculations, in line with an earlier CDF result~\cite{Aaltonen:2009iz,Ahrens:2010mj}.

\begin{figure}
\centering
\includegraphics[width=0.9\linewidth]{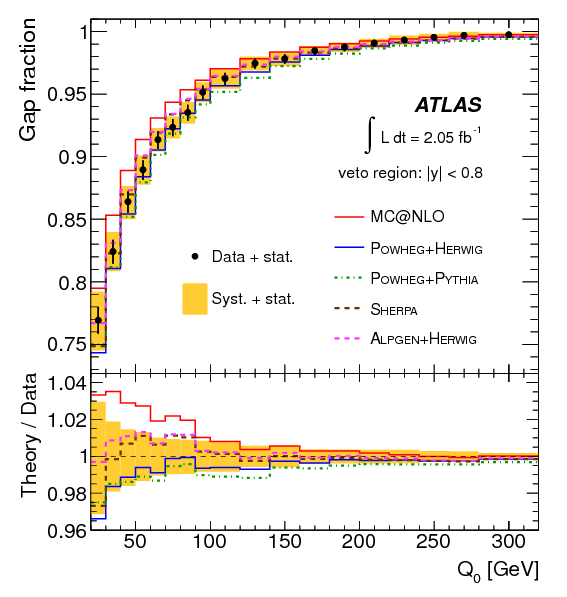}
\caption{ATLAS measurement of the fraction of $\ttbar$ events with no extra jet with $p_T>Q_0$ in the central region $|y|<0.8$, compared with various MC models~\cite{ATLAS:2012al}.}
\label{fig:gapfrac}
\end{figure}

The fraction of $\ttbar$ events with no additional jet above a given $p_T$ cut is an observable which is sensitive to differences in the modeling of QCD radiation / parton shower in the event. Both ATLAS~\cite{ATLAS:2012al} and CMS~\cite{CMS-PAS-TOP-12-023} have performed measurements of this quantity as function of the $p_T$ cut in different kinematic regimes. The results were compared with various LO or NLO MC generators, interfaced with parton showers (see Fig.~\ref{fig:gapfrac}). In the central rapidity region ($|y|<0.8$), MC@NLO overestimates the gap fraction, corresponding to too few jets being produced. Other MC models such as POWHEG, SHERPA or ALPGEN are in better agreement with the data. Comparisons with models where parameters which control the QCD radiation, such as ISR/FSR settings, the $Q^2$ scale or the matching threshold between the matrix element and the parton shower, are varied, were also performed. They show that the data are typically more precise than the spread between the assumed parameter variations. This implies that the results can be used to reduce these variations when evaluating modeling uncertainties in future measurements.

\section{$\mathbf\ttbar$+jets}

Additional jets in $\ttbar$ events can be produced through higher order QCD diagrams. Such contributions are modeled in the current MC generators either though explicit matrix elements or as a parton shower approximation. Full NLO QCD calculations also exist for $\ttbar$+1 and 2 jets.

\begin{figure}
\centering
\includegraphics[width=0.9\linewidth]{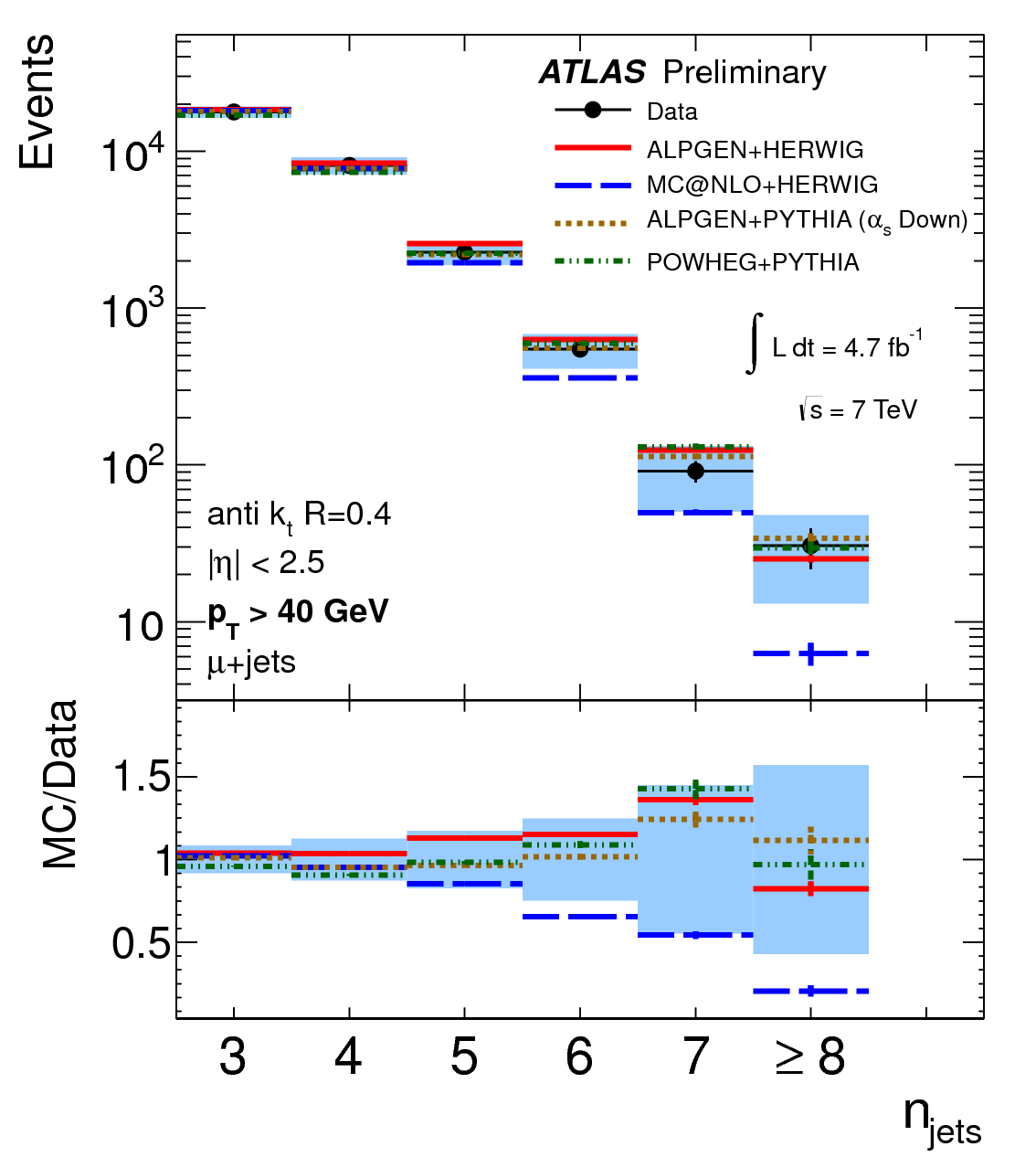}
\caption{Measurement of the jet multiplicity in $\ttbar$ events in the lepton+jets channel by ATLAS~\cite{ATLAS-CONF-2012-155}.}
\label{fig:ttjets1}
\end{figure}

\begin{figure}
\centering
\includegraphics[width=0.8\linewidth]{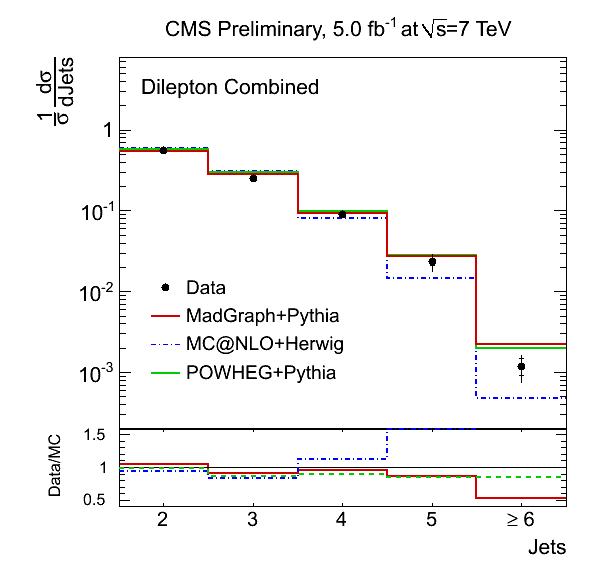}
\caption{Measurement of the jet multiplicity in $\ttbar$ events in the di-lepton channel by CMS~\cite{CMS-PAS-TOP-12-023}.}
\label{fig:ttjets2}
\end{figure}

ATLAS has measured the jet multiplicity in $\ttbar$ events in the lepton+jets channel~\cite{ATLAS-CONF-2012-155}, while CMS employed the di-lepton channel~\cite{CMS-PAS-TOP-12-023}, both making use of the full 2011 dataset. Contributions from $\ttbar$+1,2,... extra jets are expected to contribute to events with more than 5 (3) jets in the lepton+jets (di-lepton) channel, respectively. The results, shown in Figs.~\ref{fig:ttjets1} and ~\ref{fig:ttjets2}, indicate that MC@NLO underestimates large jet multiplicities, though the uncertainties are still large. This is due to the parton shower contribution being over-emphasized with respect to the matrix element. ALPGEN, POWHEG and MADGRAPH reproduce the jet multiplicity dependence better. In the future, comparisons will also be performed with the exact NLO calculations for $\ttbar$+1 and 2 jets.

\begin{figure}
\centering
\includegraphics[width=0.7\linewidth]{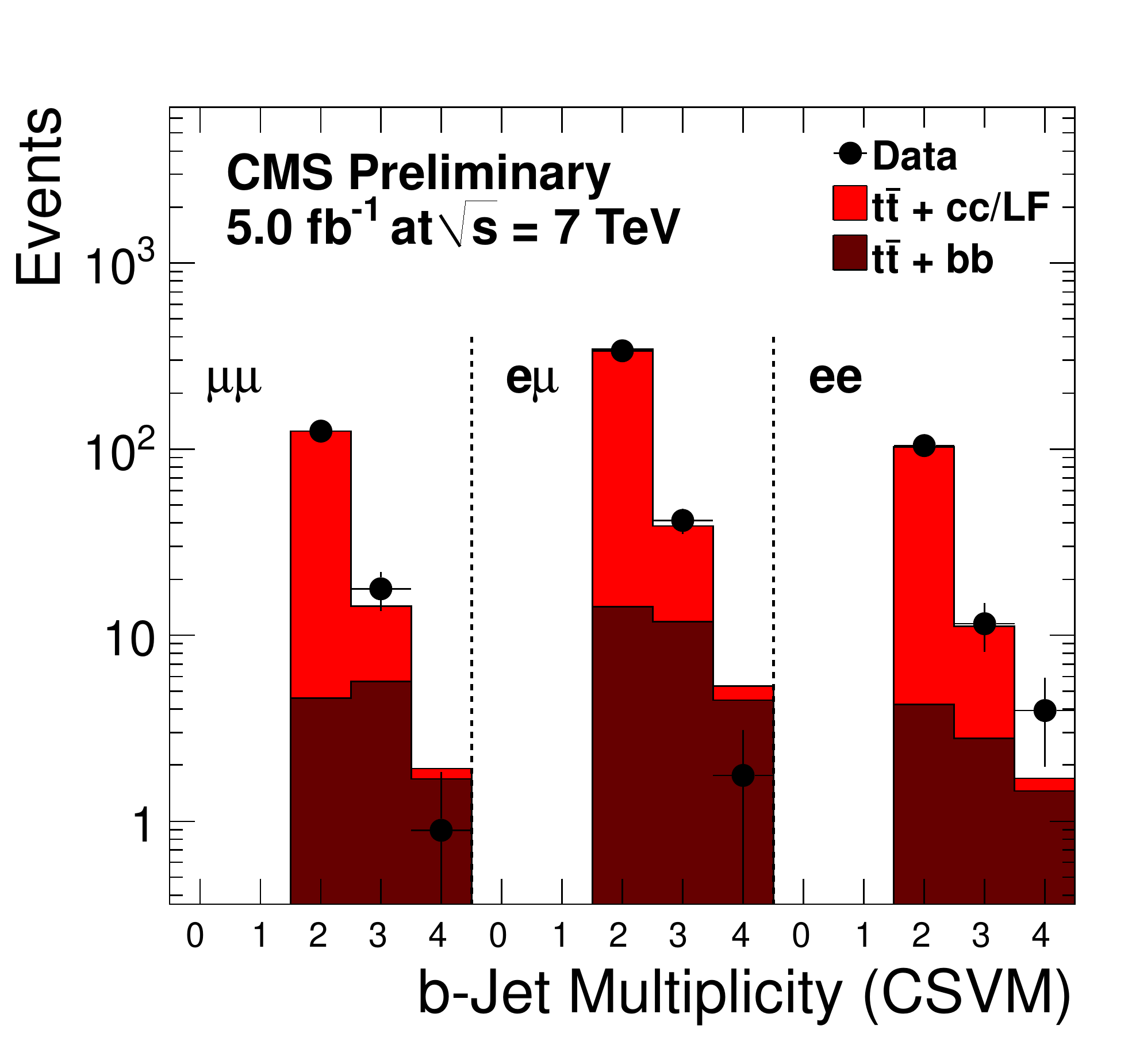}
\caption{B-jet multiplicity in di-leptonic $\ttbar$ events, used in the measurement of the $\sigma_{\ttbar b \bar{b}} / \sigma_{\ttbar j j}$  cross section ratio by CMS~\cite{CMS-PAS-TOP-12-024}.}
\label{fig:ttbb}
\end{figure}

The production of $\ttbar$ in association with two extra jets, in particular the case where they are b-jets, constitutes an important background to Higgs boson production in association with $\ttbar$, where the Higgs decays into $b\bar{b}$. CMS has performed the first measurement of the cross section ratio $\sigma_{\ttbar b \bar{b}} / \sigma_{\ttbar j j}$ in the di-lepton channel~\cite{CMS-PAS-TOP-12-024} (Fig.~\ref{fig:ttbb}). The measurement was performed by means of a fit to the b-jet multiplicity distribution and the result, quoted at particle level within the visible phase space, is  $\sigma_{\ttbar b \bar{b}} / \sigma_{\ttbar j j} = 3.6 \pm 1.1 \stat \pm 0.9 \syst \%$, to be compared with the predictions from MADGRAPH of 1.2\% and POWHEG of 1.3\%. Comparisons with  NLO QCD  calculations are pending.

\section{$\mathbf\ttbar$+boson}

The ultimate goal of measurements of the production of $\ttbar$ in association with a vector boson V (V$=\gamma,W,Z$) is to study the couplings of the top quark with bosons. Due to the small cross sections involved, these measurements are just at the beginning. 

ATLAS has performed a measurement of the cross section for the production of $\ttbar + \gamma$~\cite{ATLAS-CONF-2011-153}. The result, corresponding to the requirement $p_{T,\gamma}> 8 \GeV$, is $\sigma_{\ttbar+\gamma} \cdot \BR = 2.0 \pm 0.5 \stat \pm 0.7 \syst \pm 0.08 \lumi \pb$, consistent with NLO QCD calculations.

ATLAS has also searched for events where $\ttbar$ is produced in association with a Z-boson~\cite{ATLAS-CONF-2012-126}. In $\Lint = 4.7\ \fbinv$ of data, one candidate is observed with three leptons and two b-jets, consistent with expectations. An upper limit $\sigma_{\ttbar+Z} < 0.71 \pb$ is derived, consistent with the signal cross section at NLO of $0.14 \pb$.

\begin{figure}
\centering
\includegraphics[width=0.8\linewidth]{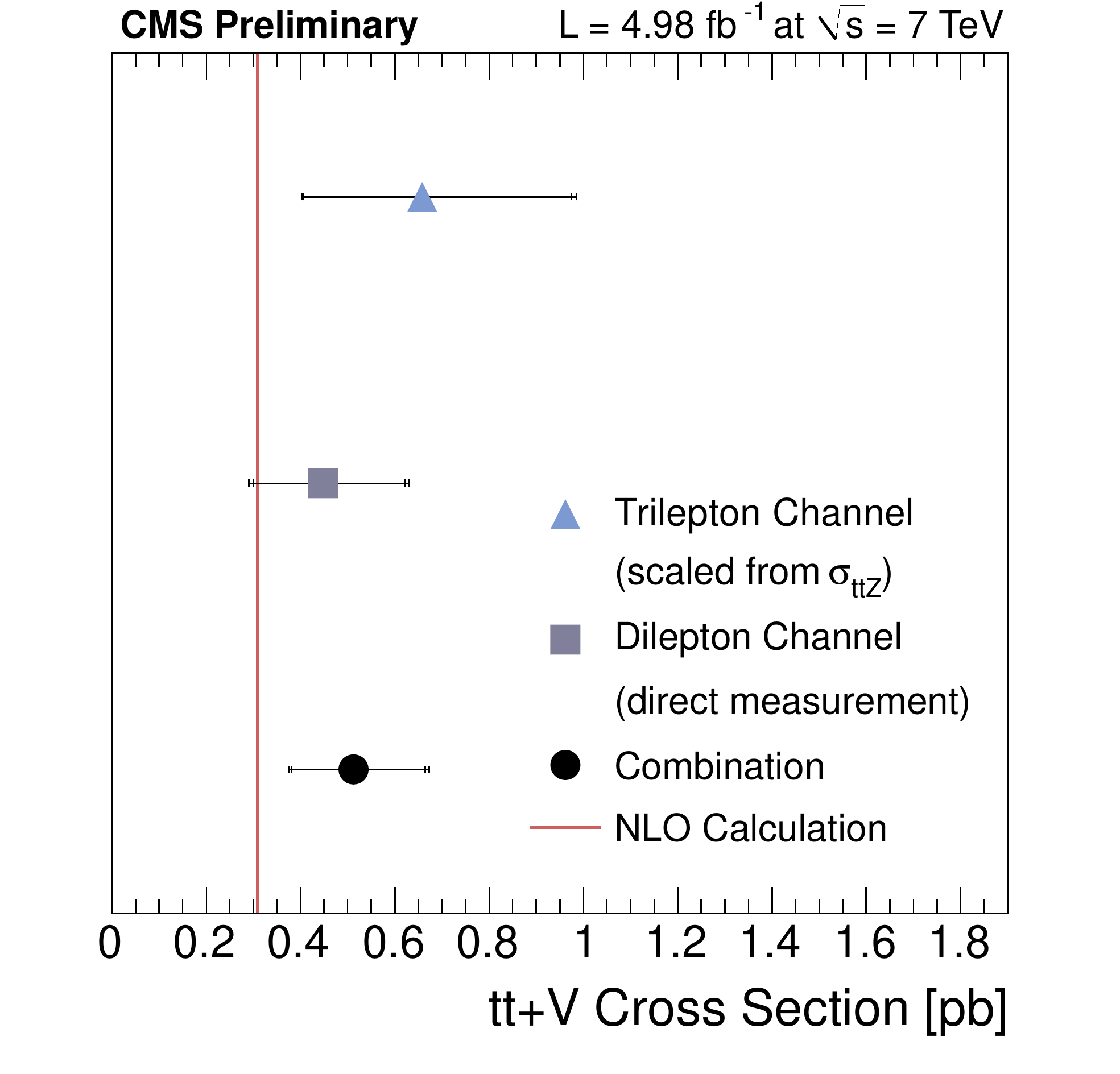}
\caption{CMS cross section measurements for the production of $\ttbar$+V in the di-lepton and tri-lepton channels as well as their combination, compared with the NLO QCD calculation~\cite{CMS-PAS-TOP-12-014}.}
\label{fig:ttv}
\end{figure}

CMS has performed a search for the production of $\ttbar$+W/Z in $\Lint = 5.0 \fbinv$ of data, employing both the di-lepton and tri-lepton final states~\cite{CMS-PAS-TOP-12-014}. The observed event numbers are consistent with a contribution from $\ttbar$+W/Z production on top of the background and cross sections were measured (Fig.~\ref{fig:ttv}). Combining the two channels, a $4.7\sigma$ significance is obtained for the production of $\ttbar$+V (V= W or Z). 

\section{Conclusions}

The era of precision top quark physics, which started at the Tevatron, is now continuing at the LHC. Experimental measurements of the total $\ttbar$ cross section have reached the precision of $5\%$, challenging the theory predictions, which are near-complete at NNLO. The first round of differential cross section measurements has been performed, which can be used not only to constrain the MC modeling of $\ttbar$ production, but also to understand this important background to Higgs production as well as in new physics searches.

The future will see even more precise total and differential cross section measurements, which can be compared with (N)NLO theory. Top quark cross-section measurements can also be used to place constraints on $m_t$, $\alpha_s$ or the gluon PDF. 
But besides precision QCD, the goal is to understand $\ttbar$ production as accurately as possible in oder to look for deviations which may be due to contributions from new physics.

\begin{acknowledgement}
The author is indebted to the ATLAS, CDF, CMS and D0 top quark physics working groups for producing the results summarized in this article.
\end{acknowledgement}

%%%%%%%%%%%%%%%%%%%%%%%%%%%%%%%%%%%%%%%%%%%%%%%%%%%%%%%%%%%%%%%%%%%%%%%%%%%%%%%

% BibTeX or Biber users please use (the style is already called in the class, ensure that the "woc.bst" style is in your local directory)

\bibliography{proc}

\begin{thebibliography}{47}

\bibitem{Deliot:2010ey}
F.~Deliot, D.A. Glenzinski, Rev.Mod.Phys. \textbf{84}, 211 (2012),
  \texttt{1010.1202}

\bibitem{topreview}
F.P. Schilling, Int.J.Mod.Phys. \textbf{A27}, 1230016 (2012),
  \texttt{1206.4484}

\bibitem{talk-palencia}
E.~Palencia, \emph{these proceedings}

\bibitem{chiarelli-talk}
G.~Chiarelli, \emph{these proceedings}

\bibitem{shabalina-talk}
E.~Shabalina, \emph{these proceedings}

\bibitem{stelzer-talk}
B.~Stelzer, \emph{these proceedings}

\bibitem{demina-talk}
R.~Demina, \emph{these proceedings}

\bibitem{talk-mitov}
A.~Mitov, \emph{these proceedings}

\bibitem{kamenik-talk}
J.~Kamenik, \emph{these proceedings}

\bibitem{Baernreuther:2012ws}
P.~Baernreuther, M.~Czakon, A.~Mitov, Phys.Rev.Lett. \textbf{109}, 132001
  (2012), \texttt{1204.5201}

\bibitem{Abazov:2011mi}
V.M. Abazov et~al. (D0 Coll.), Phys.Rev. \textbf{D84}, 012008 (2011),
  \texttt{1101.0124}

\bibitem{Aaltonen:2010ic}
T.~Aaltonen et~al. (CDF Coll.), Phys.Rev.Lett. \textbf{105}, 012001 (2010),
  \texttt{1004.3224}

\bibitem{cdf10878}
{CDF Coll.}, \emph{{M}easurement of {T}op {D}ilepton {C}ross {S}ection with
  {CDF} {F}ull data the {DIL} {S}election}, CDF Note 10878 (2012)

\bibitem{cdf10562}
{CDF Coll.}, \emph{{M}easurements of {T}op {Q}uark {P}roperties in the
  $\tau$+jets decay channel at {CDF}}, CDF Note 10562 (2011)

\bibitem{tevxscomb}
{Tevatron Electroweak Working Group}, \emph{{C}ombination of the $t\bar{t}$
  production cross section measurements from the {T}evatron {C}ollider}, D0
  Note 6363 and CDF Note 10926 (2012)

\bibitem{ATLAS-CONF-2011-121}
{ATLAS Coll.}, \emph{{M}easurement of the $t\bar{t}$ production cross-section
  in pp collisions at $\sqrt{s}=7 \rm\ {T}e{V}$ using kinematic information of
  lepton+jets events}, ATLAS-CONF-2011-121 (2011)

\bibitem{:2012cj}
S.~Chatrchyan et~al. (CMS Coll.) (2012), \texttt{1212.6682}

\bibitem{ATLAS-CONF-2012-131}
{ATLAS Coll.}, \emph{{M}easurement of the top quark pair production cross
  section with {ATLAS} in pp collisions at $\sqrt{s} = 7 \rm\ {T}e{V}$ in the
  single-lepton channel using semileptonic b decays}, ATLAS-CONF-2012-131
  (2012)

\bibitem{:2012bta}
S.~Chatrchyan et~al. (CMS Coll.), JHEP \textbf{1211}, 067 (2012),
  \texttt{1208.2671}

\bibitem{cmspas-top-11-004}
{CMS Coll.}, \emph{{M}easurement of the $t\bar{t}$ production cross section in
  the $\tau$+jets channel in $pp$ collisions at $\sqrt{s}=7$ {T}e{V}},
  CMS-PAS-TOP-11-004 (2012)

\bibitem{Aad:2012vna}
G.~Aad et~al. (ATLAS Coll.) (2012), \texttt{1211.7205}

\bibitem{ATLAS-CONF-2012-031}
{ATLAS Coll.}, \emph{{M}easurement of the $t\bar{t}$ production cross section
  in the all-hadronic channel in $4.7 \rm\ fb^{-1}$ of pp collisions at
  $\sqrt{s} = 7 \rm\ {T}e{V}$ with the {ATLAS} detector}, ATLAS-CONF-2012-031
  (2012)

\bibitem{CMS-PAS-TOP-11-007}
{CMS Coll.}, \emph{{M}easurement of the $t\bar{t}$ production cross section in
  the fully hadronic decay channel in pp collisions at 7 {T}e{V}},
  CMS-PAS-TOP-11-007 (2011)

\bibitem{lhc7combi}
{ATLAS+CMS Coll.}, \emph{{C}ombination of {ATLAS} and {CMS} top-quark pair
  cross section measurements using up to $1.1 \rm\ fb^{-1}$ of data at 7
  {T}e{V}}, ATLAS-CONF-2012-134, CMS-PAS-TOP-12-003 (2012)

\bibitem{CMS-PAS-TOP-12-022}
{CMS Coll.}, \emph{{F}irst {D}etermination of the {S}trong {C}oupling
  {C}onstant from the $t\bar{t}$ {C}ross {S}ection}, CMS-PAS-TOP-12-022 (2012)

\bibitem{Aliev:2010zk}
M.~Aliev, H.~Lacker, U.~Langenfeld, S.~Moch, P.~Uwer et~al.,
  Comput.Phys.Commun. \textbf{182}, 1034 (2011), \texttt{1007.1327}

\bibitem{Moch:2012mk}
S.~Moch, P.~Uwer, A.~Vogt, Phys.Lett. \textbf{B714}, 48 (2012),
  \texttt{1203.6282}

\bibitem{Czakon:2011xx}
M.~Czakon, A.~Mitov (2011), \texttt{1112.5675}

\bibitem{Czakon:2012zr}
M.~Czakon, A.~Mitov, JHEP \textbf{1212}, 054 (2012), \texttt{1207.0236}

\bibitem{CMS-PAS-TOP-12-006}
{CMS Coll.}, \emph{{T}op pair cross section in $e/\mu$+jets at 8 {T}e{V}},
  CMS-PAS-TOP-12-006 (2012)

\bibitem{CMS-PAS-TOP-12-007}
{CMS Coll.}, \emph{{M}easurement of the $t\bar{t}$ production cross section in
  the dilepton channel in pp collisions at $\sqrt{s}$=8 {T}e{V}},
  CMS-PAS-TOP-12-007 (2012)

\bibitem{ATLAS-CONF-2012-149}
{ATLAS Coll.}, \emph{{M}easurement of the top quark pair production cross
  section in the single-lepton channel with {ATLAS} in proton-proton collisions
  at 8 {T}e{V} using kinematic fits with b-tagging}, ATLAS-CONF-2012-149 (2012)

\bibitem{Czakon:2012pz}
M.~Czakon, A.~Mitov, JHEP \textbf{1301}, 080 (2013), \texttt{1210.6832}

\bibitem{:2012qka}
S.~Chatrchyan et~al. (CMS Coll.) (2012), \texttt{1211.2220}

\bibitem{Kidonakis:2010dk}
N.~Kidonakis, Phys.Rev. \textbf{D82}, 114030 (2010), \texttt{1009.4935}

\bibitem{Abazov:2010js}
V.~Abazov et~al. (D0 Coll.), Phys.Lett. \textbf{B693}, 515 (2010),
  \texttt{1001.1900}

\bibitem{CMS-PAS-TOP-12-019}
{CMS Coll.}, \emph{{M}easurement of missing transverse energy in top pair
  events}, CMS-PAS-TOP-12-019 (2012)

\bibitem{Aad:2012hg}
G.~Aad et~al. (ATLAS Coll.) (2012), \texttt{1207.5644}

\bibitem{Aaltonen:2009iz}
T.~Aaltonen et~al. (CDF Coll.), Phys.Rev.Lett. \textbf{102}, 222003 (2009),
  \texttt{0903.2850}

\bibitem{Ahrens:2010mj}
V.~Ahrens, A.~Ferroglia, M.~Neubert, B.~Pecjak, L.~Yang, Nucl.Phys.Proc.Suppl.
  \textbf{205-206}, 48 (2010), \texttt{1006.4682}

\bibitem{ATLAS:2012al}
G.~Aad et~al. (ATLAS Coll.), Eur.Phys.J. \textbf{C72}, 2043 (2012),
  \texttt{1203.5015}

\bibitem{CMS-PAS-TOP-12-023}
{CMS Coll.}, \emph{{M}easurement of {J}et {M}ultiplicity {D}istributions in
  {T}op {Q}uark {E}vents {W}ith {T}wo {L}eptons in the {F}inal {S}tate at a
  centre-of-mass energy of 7 {T}e{V}}, CMS-PAS-TOP-12-023

\bibitem{ATLAS-CONF-2012-155}
{ATLAS Coll.}, \emph{{M}easurement of the jet multiplicity in top- anti-top
  final states produced in 7 {T}e{V} proton-proton collisions with the {ATLAS}
  detector}, ATLAS-CONF-2012-155 (2012)

\bibitem{CMS-PAS-TOP-12-024}
{CMS Coll.}, \emph{{F}irst {M}easurement of the {C}ross {S}ection {R}atio
  $\sigma(t\bar{t}b\bar{b})/\sigma(t\bar{t}jj)$ in pp {C}ollisions at
  $\sqrt{s}$=7 {T}e{V}}, CMS-PAS-TOP-12-024 (2012)

\bibitem{ATLAS-CONF-2011-153}
{ATLAS Coll.}, \emph{{M}easurement of the inclusive $t\bar{t}\gamma$ cross
  section with the {ATLAS} detector}, ATLAS-CONF-2011-153 (2011)

\bibitem{ATLAS-CONF-2012-126}
{ATLAS Coll.}, \emph{{S}earch for $t\bar{t}$+{Z} production in the three lepton
  final state with $4.7 \rm\ fb^{-1}$ of $\sqrt{s}$= 7 {T}e{V} pp collision
  data collected by the {ATLAS} detector}, ATLAS-CONF-2012-126 (2012)

\bibitem{CMS-PAS-TOP-12-014}
{CMS Coll.}, \emph{{F}irst {M}easurement of {V}ector {B}oson {P}roduction
  {A}ssociated with {T}op-{A}ntitop {P}airs at 7 {T}e{V}}, CMS-PAS-TOP-12-014
  (2012)

\end{thebibliography}

\end{document}